\documentclass[a4paper,USenglish,cleveref, autoref, thm-restate]{lipics-v2021} 

\usepackage{color}
\usepackage{colortbl}
\usepackage{multicol}
\usepackage{booktabs, tabularx, makecell, multirow,caption}
\usepackage{subcaption}
\usepackage{amsmath}
\usepackage{amsthm}
\usepackage{balance}
\usepackage[ruled, noline, linesnumbered, commentsnumbered, noend]{algorithm2e}
\usepackage[end]{algpseudocode}
\usepackage{xspace}
\usepackage{booktabs, tabularx, makecell, multirow}
\usepackage{thmtools}
\usepackage{graphicx}
\usepackage{subcaption}
\usepackage{todonotes}
\usepackage[sort]{cite}

\newtheorem{mydefinition}{Definition}
\newtheorem{mytheorem}{Theorem}
\newtheorem{myLemma}{Lemma}
\newtheorem{myInvariant}{Invariant}

\newtheorem{myAssumption}{Assumption}

\newcommand{\system}{\emph{H\&A}\xspace}

\newcommand{\myHash}[1]{hash(#1)}
\newcommand{\mymyHash}{hash}

\newcommand{\MSCFS}{MSCFS\xspace}
\newcommand{\MSCFSs}{MSCFSs\xspace}

\newcommand{\ServerList}{ServerList\xspace}
\newcommand{\ServerLists}{ServerLists\xspace}

\newcommand{\hosts}{heads\xspace}

\newcommand{\head}[1]{head(#1)}
\newcommand{\hostt}[1]{host^t(#1)}
\newcommand{\dis}[2]{pos_{#1}(#2)}
\newcommand{\dist}[2]{pos^t_{#1}(#2)}
\newcommand{\invt}[1]{I^t(#1)}
\newcommand{\invtotalt}[1]{I^{#1}}
\newcommand{\capUp}[1]{\lceil \frac{m^{#1}}{n^{#1}} \rceil + \alpha}

\DeclareMathOperator*{\argmin}{arg\,min}

\newcommand{\OPT}{\emph{OPT}\xspace}
\newcommand{\WBL}{\emph{WBL}\xspace}

\newcounter{myProcedure}


\SetStartEndCondition{ }{}{}
\SetKwIF{If}{ElseIf}{Else}{if}{:}{elif}{else:}{}%
\SetKwFor{While}{while}{:}{fintq}%
\SetKwFor{For}{for}{:}{fintq}%

\newcommand{\appsymb}{$\star$}
\newcommand{\appref}[1]{{\hyperref[proof:#1]{(\appsymb)}}}

\newcommand{\appendixproof}[2]{%
  \gappto{\appendixProofs}
  {
    \subsection{Proof of \cref{#1}}\label{proof:#1}
    #2
  }
}

\title{Hash \& Adjust: Competitive Demand-Aware Consistent Hashing}
\author{Arash Pourdamghani}{TU Berlin}{pourdamghani@tu-berlin.de}{0000-0002-9213-1512}{}
\author{Chen Avin}{Ben-Gurion University of the Negev}{}{0000-0002-6647-8002}{}
\author{Robert Sama}{University of Vienna}{}{0009-0002-1831-0413}{}
\author{Maryam Shiran}{TU Berlin}{}{0009-0009-9873-9172}{}
\author{Stefan Schmid}{TU Berlin \& Fraunhofer SIT}{}{0000-0002-7798-1711}{}

\authorrunning{A. Pourdamghani, C. Avin, R. Sama, M. Shiran, S. Schmid}

\Copyright{Arash Pourdamghani, Chen Avin, Robert Sama, Maryam Shiran, Stefan Schmid}

\supplement{}
\supplementdetails[subcategory={Source Code}, cite={}, swhid={}]{}{https://github.com/inet-tub/Hash-And-Adjust}

\funding{This project has received funding from the European Research Council
(ERC) under grant agreement No. 864228 (AdjustNet), 2020-2025.}



\nolinenumbers
\begin{document}
\maketitle

\begin{abstract}
Distributed systems often serve dynamic workloads and resource demands evolve over time. Such a temporal behavior
stands in contrast to the static and demand-oblivious nature of most data structures used by these systems. 
In this paper, we are particularly interested in consistent hashing, a fundamental building block in many large distributed systems. Our work is motivated by the hypothesis that a more adaptive approach to consistent hashing can leverage structure in the demand, and hence improve storage utilization and reduce access time. 

We initiate the study of demand-aware consistent hashing. Our main contribution is H\&A, a~constant-competitive online algorithm (i.e., it comes with provable performance guarantees over time). H\&A is demand-aware and optimizes its internal structure to enable faster access times, while offering a high utilization of storage. We further evaluate H\&A empirically.
\end{abstract}

\section{Introduction}
\label{sec: intro}

Access patterns and demand in distributed systems are often dynamic and feature structure over time~\cite{denning1968working,Locality99, nicopolitidis2006exploiting,Locality12,roy2015inside,sigmetrics20complexity,Locality22,SeedTree,alenex25,OPODIS25}.
Such dynamicity and temporal locality of demand have been observed in many application domains and encouraged the design of optimized computer architectures that support such structure, e.g., through a cache hierarchy. 
Our paper is motivated by the hypothesis that there is still much-untapped potential for adaptive and demand-aware approaches for optimizing the performance of data-centric applications~\cite{avin2018toward}. 

We are particularly interested in consistent hashing~\cite{KargerClassic1}, a key component of many distributed systems, which has applications in systems such as Amazon DynamoDB~\cite{Dynamo07, DynamoDB12} and Apache Cassandra~\cite{Cassandra10}.
Consistent hashing aims to provide fast lookups on top of flexible insertion and deletions, making it especially attractive for web applications and distributed databases~\cite{WebCaching99}. 
However, consistent hashing today is still mostly demand-oblivious and has shortcomings, e.g., related to the high variance in the load of different servers, which can result in sub-optimal utilization. 

This paper envisions a more demand-aware approach to consistent hashing, facilitating the adaptation to---and exploitation of---structure in the demand, with the goal of reducing access costs and improving load balancing (i.e., maximizing storage utilization).  
Our work is further motivated by a recent work by Mirrokni, Thorup and Zadimoghaddam~\cite{MirrokniTZ18} that showed that bounding server capacities within a factor of the minimum needed capacity (i.e., the number of items divided by the number of servers) is possible given a tolerable access cost increase. Considering access costs is a new dimension in the design of consistent hashing methods that we further develop in this work. 
We show that our demand-aware approach can improve storage utilization while keeping the access cost competitive.
Designing a demand-aware variant of consistent hashing is, however, challenging. This is because the demand is not perfectly known ahead of time. Therefore such a design requires a careful balancing of the benefits of adjustment and the cost of adjustments~\cite{Roughgarden19}.% That is why we need to go beyond worst-case analysis and focus on amortized analysis~\cite{}.

Our main contribution is \emph{Hash \& Adjust}
(\system for short), an online and demand-aware algorithm for consistent hashing.
The performance of \system differs only by a constant factor from the optimal offline algorithm (that knows the whole demand in advance), i.e., \system is constant competitive.
Constant competitiveness becomes possible in our system by keeping the recently accessed items close to their original server via self adjustments, in the spirit of strategies known from list update problems~\cite{SleatorT85,Albers03OnlineBook}.
Furthermore, adjustments of \system are local. Local adjustments are crucial for its practical use cases in distributed systems, due to their storage and computation efficiency.
To achieve this, we extend list access in a novel direction: \emph{list access with multiple heads and with capacity}. This extension can be of independent interest in other application scenarios. 

We complement our theoretical contribution with an empirical comparison between the performance of \system with state-of-the-art algorithms.
Our empirical evaluations indicate an average $54\%$ improvement in the access cost (i.e., item access time) compared to the recent consistent hashing method~\cite{MirrokniTZ18} (see details in the ~\Cref{sec: implement}).

\section{Motivation and Overview of \system}
\label{sec: motivate}

This section motivates the need to critically reconsider today's consistent hashing methods. We begin with a short description of the current state of consistent hashing methods, detailed in more depth in Section~\ref{sec: related}.

\noindent \textbf{Basic consistent hashing.}
In the basic version of consistent hashing~\cite{KargerClassic1}, servers (e.g., nodes of a system) are arranged over a ring, hosting a set of items (e.g., DNS records). 
Operations on an item begin by computing the hash value for the item. Then, based on the hash value, the item is assigned to the closest server (in a clockwise manner, details follow in the next section).
In practice, the requests are generated in a distributed manner: any server can initiate the request, run the hash function, and find the target server using the network that is built on top of the ring.
In this work, following the lead of recent advancements~\cite{MirrokniTZ18, AamandKT21}, we focus on ring abstraction. With this abstraction, we provide flexibility of choice: 
we can connect or even better: nodes on the ring can be connected to form a desired network of choice. Hence, we remove the dependence on a particular network and analyze a consistent hashing instance in its most general form. 

Consistent hashing operations are  \emph{history independent} and \emph{local}. This is because a single shared hash function suffices to determine the server that hosts an item. 
Furthermore, consistent hashing is preferable to other distributed hashing methods as it allows for online insertion and deletions.
However, traditional consistent hashing methods suffer from low storage utilization. Therefore, we also evaluate server loads (i.e., the number of items in a server).

\noindent \textbf{Suboptimal storage utilization.}
A high variance in server loads implies a suboptimal storage utilization and is generally undesirable in practice. 
To this end, we define server capacity as the maximum number of items a server can hold.
Since the server with the maximum load can not be predetermined, we need to reserve the capacity equal to the maximum load for \emph{all} servers.
As the sum of the load of all servers is equal to the number of items, we define the storage utilization as:
\begin{equation}
    \label{eg: load}
   \frac{\text{Total Load}}{\text{Total Capacity}} = \frac{\text{Number of Items}}{\text{Maximum Load * Number of Servers}}  = \frac{\text{Average Load}}{\text{Maximum Load}}
\end{equation}
For a fair comparison, we consider an equal number of items and servers for all algorithms. Hence, given Equation~\ref{eg: load}, we focus on the maximum load of algorithms.

\begin{table}[t]
\small
\begin{center}
\begin{tabular}{| c | c |c | }
    \hline
   Algorithm &  \cellcolor{gray!20} Access Cost
   &  Memory Utilization
   \\ \hline
    \cellcolor{gray!20}
    Traditional~\cite{KargerClassic1}
    & \cellcolor{green!20} Low
    & \cellcolor{red!20} Low 
    \\ \hline
    WBL~\cite{MirrokniTZ18}
    & \cellcolor{red!20} High 
    & \cellcolor{yellow!20} Medium 
   \\ \hline
    \cellcolor{gray!20}
    \system [this work]
    & \cellcolor{green!20} Low                      
    & \cellcolor{green!20} High \\
 \hline
 
\end{tabular}
\caption{A comparison between \system and other variants of consistent hashing, in terms of access cost and memory utilization. \system ensures both high memory utilization and low access cost by supporting bounded server loads and on-the-fly self-adjustments, while other algorithms can not support both at the same time. The empirical counterpart of this table can be observed in Figure~\ref{fig: intro-prac} in Section~\ref{sec: implement}.}
\label{fig: intro}
\end{center}
\end{table}

% \begin{figure}
%     \centering
% \includegraphics[width=0.2\linewidth, clip, trim={8 1 5 10}]{Plots/1.pdf}
%     \caption{Figure shows the load distribution of algorithms mentioned in Table~\ref{fig: intro}. ``Max." stands for maximum load. We assume around $8500$ items to $200$ servers. 
%     }
%     \label{fig: loads-per-server}
% \end{figure}

It is well known that traditional consistent hashing has a non-constant difference between the average and maximum load~\cite{harvey2022book}. 
This is why a recent work by Mirrokni~et~al.~\cite{MirrokniTZ18} suggested bounding the load of each server by a constant factor times the minimum load needed (we call their work With Bounded Loads, or ``WBL" for short). 
However, the difference remains limited between \WBL's proposal
and the case with unbounded capacity %(see Figure~\ref{fig: loads-per-server} for an example of the load distribution of algorithms from Table~\ref{fig: intro}). 
We emphasize that bounding loads aim to adhere to stringent storage constraints of reserved storage resources, that can not be achieved through flexible storage allocation methods~\cite{on-demand1}.

\noindent \textbf{Increased access time.}
The idea of bounding server capacity might seem obvious at first sight. However, a challenge arises: as servers become full, we need to put the items of the oversubscribed server elsewhere, otherwise data may be lost.
To model this issue more formally, we define access time as the number of servers that one needs to traverse to find an item. This is a fair consideration, as moving items between servers requires noticeably more time than searching within a server for an item. 
Also,
moving items between servers
should be local: without a centralized map of items or separate server assignment functions.

Fortunately, however, real-world request sequences have an inherent \emph{temporal locality}.  This means that among items that are assigned to a server, we have intervals of consecutive requests for the same item -- see~\cite{sigmetrics20complexity} for a study of the temporal locality of real-world instances.

\noindent \textbf{Overview of \system.}
The consistent hashing method proposed in this work relies on an online algorithm, \system. 
\system allows us to solve the two issues identified above (and summarized in Table~\ref{fig: intro}).
\begin{itemize}
    \item Our method improves \emph{storage utilization} by considering only an \emph{additive} additional capacity (e.g., considering only $2$ more storage slots per server) rather than the multiplicative additional capacity proposed previously. Thus, we can push the storage utilization to its limit.
    \item To reduce \emph{access time}, \system self-adjusts items between servers after every request -- not only upon insertion or deletion. This way, we ensure a decreased overall average cost, especially in the case of a high temporal locality.
\end{itemize}

\section{Model and Preliminaries}
\label{sec: model}
In this section, we present our theoretical model and introduce the required terminologies and preliminaries.

\noindent \textbf{Consistent hashing.}
We consider a set of $m$ \emph{items} $V$, and a set $S$ of $n$ \emph{servers}.
As in practice, the number of items is much larger than the number of servers; we consider $m$ to be significantly greater than $n$, i.e., $m \gg n$.

We assign a \emph{head} to each item. The head of an item is a server that has the closest hash value to the hash value of the item.
In our analysis, we consider a hash function $\myHash$, that hashes the ID of a server or an item to a value in the range $[0,1]$ uniformly at random\footnote{This is a common~\cite{MirrokniTZ18} assumption, given that there are hash functions that can provide almost uniformly at random results~\cite{5Independent1}.}.  
Formally, the head $s$ for an item $v$ is selected as follows:
\[
\head{v} = \argmin_{s \in S} \ \myHash{s}-\myHash{v} \mod 1
\]
In other words, the function $\mymyHash$ maps items and servers on a ring, and the head of an item is a server on the ring closest to the item in a clockwise manner.
We emphasize that this process is carried out in a distributed fashion, i.e., the hash function is known globally, not solely to a~central entity.
Furthermore, we define the server that hosts item $v$ at time $t$ as $\hostt{v}$.

Considering a fixed $\mymyHash$ function, servers maintain a~certain order between them. Given a server $s$, we denote the server after server $s$ by $s^+$ and the server before $s$ by $s^-$.
%An item can move from a server to its next and previous server locally at a cost of $1$.

\noindent \textbf{Bounding servers' capacity.}
%Traditional consistent hashing methods do not consider a bound on the capacity of servers. This results in unequal distribution of items between servers.
%By bounding the capacity of all servers and moving items to nearby servers~\cite{MirrokniTZ18}, we can guarantee a much better load distribution among servers (as discussed in the motivation section above).
All servers can contain up to $c = \lceil \frac{m}{n} \rceil + \alpha$ items, i.e., they have \emph{capacity} equal to $c$.
Variable $\alpha \ge 1$ is the \emph{additive extra capacity} which is fixed throughout the run time of \system. See Figure~\ref{fig: model} for an example.

Considering additive extra capacity results in a tradeoff between storage utilization and access cost. 
Increasing the value of $\alpha$ would lower the access cost of the system (as items will be stored closer to their head), but requires more storage per server. 

\noindent \textbf{Online requests sequence.}
We consider a significantly large request sequence $\sigma = (\sigma_1,\dots)$, in which $\sigma_t$ is the request at time $t$.
Requests can either refer to an item or a server, and they might have one of three types: 
\begin{itemize}
    \item \emph{Access} to an item: searching for an item inside the system,
    \item \emph{Inserting} a server or an item: adding a new server or an item that was not in the system before,
    \item \emph{Deleting} a server or an item: removing one of the existing items or servers from the system.
\end{itemize}

It is important to respond to requests locally and immediately, i.e., not requiring a global view or a (estimation of) future knowledge when updating the data structure, as they are critical for distributed settings. Furthermore, we assume that insertion and deletion operations are happening rarely~\cite{RevisitCH21}, in particular after every $\sum_{i=1}^{n-1} e^{-\frac{\alpha^2}{2m}(i+1)}$ access operation.
While responding to requests, a system might \emph{reconfigure} some items, i.e., moving them from one server to another.

\noindent \textbf{Cost model.}
Responding to a request comes at a delay:
we might need to search a few servers to find the item or move items around after insertion and deletion requests. 
Therefore, we consider two types of delay:
\begin{itemize}
    \item \emph{Search delay:} the delay that occurs while searching and retrieving data from a server. In our abstraction, we consider the access delay as the number of servers traversed to find an item, including the head of an item.
    \item \emph{Reconfiguration delay:} the delay of moving items between servers. Similar to the access delay, we only consider the number of servers that we move an item between.
\end{itemize}

We emphasize that all the delays in any operation (access, insertion, and deletion) are essentially either a search delay or a reconfiguration delay.
We abstract the delay we observe for a request as the \emph{cost} of that request. The cost of searching or moving items inside a server is negligible compared to the inter-server costs. We point out, that in our model, accessing the head of an item has a cost of $1$, for any algorithm.

Given that transferring data costs more than accessing it,
we consider a constant factor $\omega > 1$ for the cost of relocating a single item from a server to a neighboring server compared to the cost of going from one server to another searching for an item
(this is another extension of the previous consistent hashing models~\cite{MirrokniTZ18, AamandKT21}, which only considered $\omega=1$).
Hence, the \emph{total} cost of an algorithm $ALG$ for all its operations over a request sequence is:
\begin{equation}
\label{eq: total cost}
    C_{ALG}(\sigma) = C^{\text{Search}}_{ALG}(\sigma) + \omega \cdot C^{\text{Reconfiguration}}_{ALG}(\sigma) 
\end{equation}

Our aim is to minimize the total cost of our online algorithm compared to the optimal offline algorithm.
We consider that the inputs are generated by an oblivious adversary. The oblivious adversary is not aware of the random bits used by an algorithm. This is a common assumption, as there are opportunities for pseudorandom number generation (for details on different types of adversaries, check~\cite{borodin2005onlineBook}).  
Formally, our objective is to develop an algorithm with a constant competitive ratio, i.e., we want the cost of our algorithm to match the optimal offline algorithm asymptotically. 

\begin{mydefinition}[Competitive ratio]
Consider an online algorithm $ALG$, and an optimal offline algorithm $OPT$. Denote the total cost of $ALG$ over an input sequence $\sigma$ as $C_{ALG}(\sigma)$, and similarly the cost of $OPT$ as $C_{OPT}(\sigma)$. Then the (strict) competitive ratio of $ALG$ is defined as:
$\max_{\sigma} \frac{C_{ALG}(\sigma)}{C_{OPT}(\sigma)}$.
\end{mydefinition}

\section{\system Algorithms}
In this section, we first discuss how access operations are done in \system, and then discuss insertion and deletion operations.

\subsection{Access in \system}
We start by describing how we handle an access request in \system. The main innovation of our approach is to use self-adjustments after each access request.

\noindent \textbf{Self-adjustments on a ring.}
With self-adjustments, we aim to reduce the access cost by adjusting the position of items inside the data structure. In particular, we look at a self-adjustment procedure that brings back items to their heads step by step, from a server to its neighbor on the ring. 
We consider self-adjustments on a ring as an abstraction (rather than considering a fixed network used in distributed hash tables). This is a plus: it provides a flexible choice for the network that can be built on top, as each application in peer-to-peer usecases uses a different overlay network (Chord, Kademlia, etc.).

\noindent \textbf{Accessing an item.}
After our algorithm (Algorithm~\ref{Alg: MTHC}) receives access to an item $v$, it starts looking for $v$ inside the head assigned to it, $s = \head{v}$.
From then on, for any given server $s$, \system checks for the following three possibilities in order: 
\begin{enumerate}
    \item Item $v$ is not in server $s$ and $s$ is full. Then, item $v$ might have been moved to server $s^+$, so we set $s = s^+$ and start checking again.
    \item Item $v$ is not in server $s$ and $s$ is not full. This means that item $v$ is not in our system.
    \item Item $v$ is in server $s$. Then we successfully found the item. If $s \neq \head{v}$, then we swap $v$ with the least recently accessed item in server $s^-$, $u$, through a $swap(u,v)$ operation.
    The swap operation for items $u$ and $v$ works for two items with adjacent servers.
    We repeat swapping until $v$ reaches  $\head{v}$. 
\end{enumerate}

\begin{algorithm}[t]
\caption{\system access to an item $v$}
\label{Alg: MTHC}
\small
 set $s=\head{v}$. \\
\While{$v$ is not in $s$ and $s$ is full}{
    set $s$ as $s^+$.
}
\If{$v$ is not in $s$ and $s$ is not full}{
 return “$v$ is not in the system”.
}
\While{$s$ is not $\head{v}$}{
    name $u$ as the least recently accessed item in $s^-$.\\
   $swap(v,u)$. \\
    set $s$ as $s^-$.
}
\end{algorithm}

Accessing only involves items that are already in the system. If an item is not already in the system, it is handled by insertion procedure, discussed next.

\subsection{Insertion and Deletion in \system}
\label{sec: insert}
We now focus on how the insertion and deletion of items (or servers) are performed.
Now that both numbers of items and servers could change, we define $m^t$ to be the number of items and $n^t$ to be the number of servers at time~$t$. 
Given the dynamicity of $n^t$ and $m^t$, changing the servers' capacity is essential to ensure our capacity constraint on each server.

\noindent \textbf{Capacity Change.}
We perform capacity changes in phases. A phase ends in one of two scenarios:
\begin{itemize}
    \item if we have a server insertion/deletion,
    \item if the difference between the number of item insertions and deletions overshoots $n$ since the start of the phase. We keep this difference in a parameter called $\delta$.
\end{itemize}
At the end of a phase, we need to decrease or increase the capacity of all servers to adhere to the capacity constraint based on the number of items and servers.
In case of a capacity decrease, we might have some servers that have more items than their capacity. With the capacity increase, servers that were previously full are not full anymore, affecting access operations.
To remedy this issue, we first define \emph{valid} items for a server $s$.
\begin{mydefinition}
    An item is valid for server $s$ if its head is $s$ or any of the servers located in front of $s$ but before the next non-full server.
\end{mydefinition}

Given the definition of valid items, we now detail how \system solves the aforementioned issues:

\begin{enumerate}[(i)]
    \item \emph{Filling unused capacity:}
    In this case, we fill the additional slot that has been created in $s$ by moving the newest \emph{valid} item from the back of $s$ (server $s^+$ and beyond) to $s$. 
    In particular, we first search the server $s^+$ for a valid item and move it to $s$. If there is none, we look into the next servers until reaching a non-full server (given that we have extra capacity for all servers, as we have extra capacity per server, i.e, $\alpha \ge 1$, such a non-full server could always be found).
    In case no such valid item exists until the next non-full server, it means no items from the server in front of $s$ have been moved over $s$, so we keep $s$ as non-full and stop
    %(see Algorithm~\ref{proc: extra})
    , while being sure that Invariant~\ref{inv: between} is intact.
%\begin{algorithm}
%    \caption{Filling Unused Capacity}
%\label{proc: extra}

%\end{algorithm}
\begin{algorithm}[t]
\caption{\system Item Insertion/Deletion}
\label{alg: item}
\small
use Algorithm~\ref{Alg: MTHC} to search for the item $v$.\\
\If{the request is item insertion and $v$ was not found}{
Put $v$ in the first non-full server after $\head{v}$.\\
increase $\delta$ and run change capacities if $\delta \ge n$.
}
\If{the request is item deletion and $v$ was  found}{
remove $v$ from its current server.\\
fill the extra capacity.\\
decrease $\delta$ and change capacities if $\delta \le -n$.
}
\end{algorithm}
\begin{algorithm}[t]
 \caption{\system Capacity Change}
\label{alg: cap}
\small
Set the capacity of all servers equal to $\capUp{t}$.\\
Set $\delta$ equal to $0$. \\
\For{all servers $s$ starting from server $0$}{
        \If{$s$ was full in the previous phase and had extra capacity}{
         \While{Server $s$ has extra capacity}{
        set $s' = s^+$.\\
        \While{$s'$ does not have a valid item}{
        $s' = s'+1$
        }
        bring the newest valid item from $s'$ to $s$,
        }
        }
        \If{$s$ has more items than its capacity}{
            move the least recently accessed item of $s$ to $s^+$.
        }
}
\end{algorithm}

    \item \emph{Moving out extra load:}
    In this case, we need to move out the extra item $v$. For that, we move the least recently accessed item of server $s$ to server $s^+$. 
\end{enumerate}
Combining the above procedures, we end up with the algorithm to keep the capacity of servers consistent, Algorithm~\ref{alg: cap}.
Now, we discuss how insertion and deletions are designed on top of capacity change operations.

\noindent \textbf{Inserting an item.} 
To insert $v$, after searching for it, we assume that we have not found it, otherwise, we do not need to do anything, to avoid duplicates of the same item.
Then, we place $v$ in $s^{nf}$, the first non-full server after $\head{v}$. As $s^{nf}$ is non-full, we can place $v$ in it, without moving other items.

\noindent \textbf{Deleting an item.} 
To delete $v$, if we could not find it between $\head{v}$ and the next non-full server, it is nowhere else to be found (based on Invariant~\ref{inv: between}),  and nothing else is needed to be done.
If the item was found, we would remove it from the server that currently contains $v$, $host(v)$. After that, our algorithm needs to fill the extra capacity created. 
%Details of how to fill this extra capacity are in Appendix~\ref{appendix: cap change}.
If we can not find a valid item to fill the extra capacity, we do not take any further action.

\noindent \textbf{Inserting a server.} 
We first determine the position of the server using our hash function. We then tag the new server as a server that was previously full and run Algorithm~\ref{alg: cap} to bring valid items back to this server. 
Keep in mind that there might not be enough valid items, and this server can remain non-full.

\noindent \textbf{Deleting a server.} 
To delete server $s$, we move all of its items to the server $s^+$ temporally, and then remove it from the system. After running Algorithm~\ref{alg: cap} to adjust the capacity of all servers, we move the extra items that might be accumulated on top of server $s^+$ to the next servers, giving priority to the oldest items in each server to stay in that server. 

An important invariant of discussed insertion and deletion operations is that they keep the relative order of items.

\begin{figure}
    \centering
    \captionsetup[subfigure]{justification=centering}
    \begin{subfigure}[t]{0.3\textwidth}
    \centering
    \includegraphics[width=\linewidth, clip, trim= {0 10 10 0}]{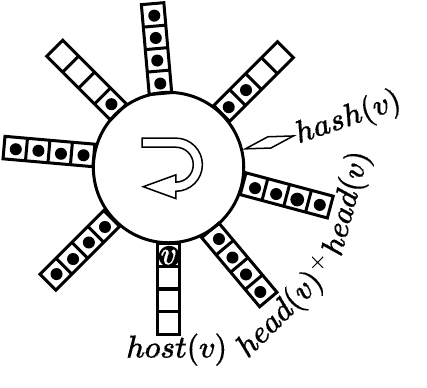}
    \caption{Consistent hasing model.}
    \label{fig: model}
    \end{subfigure}
    \hspace{20mm}
    \begin{subfigure}[t]{0.3\textwidth}
    \centering
    \includegraphics[width=0.9\linewidth, clip]{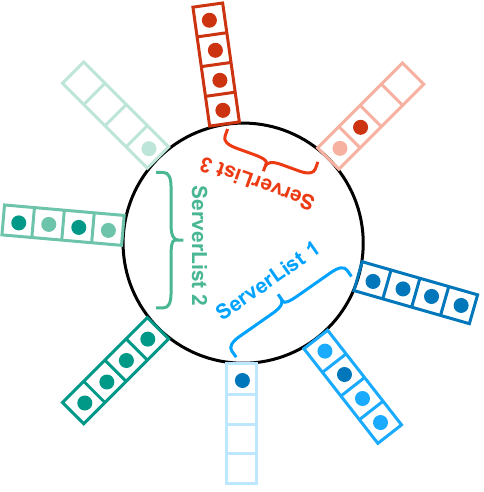}
    \caption{\ServerList decomposition.}
    \label{fig: decomposition}
    \end{subfigure}
    \caption{
    Figure~\ref{fig: model} shows an example of our model with $8$ servers, each with capacity $4$, and $24$ items. In this example, item $v$ was initially inserted into server $\head{v}$ (it had the closest hash value); however, because $\head{v}$ and $\head{v}^+$ were full, it is moved to $host(v)$.
    Figure~\ref{fig: decomposition} depicts decomposition of the previous example into \ServerLists. Each \ServerList is shown by a different color, and servers have different gradients of the color of their \ServerList as they are different \hosts on their own. Items are colored by the color of their host.
    }
\end{figure}

\begin{myInvariant}
\label{inv: order}
After an item (or a server) insertion (or deletion), our algorithm keeps the order of items that were previously in the system, the same. 
\end{myInvariant}

\section{Competitiveness}
In this section, we provide the competitiveness analysis of our algorithms. We first start by stating a general method, and then discuss the competitiveness of access and insertion/deletion operations separately.

\noindent \textbf{Decomposition to \ServerLists.}
For analytical purposes, we decompose the ring used by \system into a set of full servers, except their last server: a set of \ServerLists. Informally, a \ServerList is a consecutive subsequence of servers on the ring
(a formal and more general definition follows, see Figure~\ref{fig: decomposition} for an example).
Such a decomposition is possible because of additional extra capacity, which results in non-full servers.
Those non-full servers are at the end of a \ServerList, and the server after them is the start of another \ServerList. 
%Hence, decomposition is possible at each point in time.
Our algorithm is designed such that no item moves over the non-full servers; hence the interactions inside a \ServerList do not affect other \ServerLists.

\begin{myInvariant}
\label{inv: between}
For any given item $v$ at any given time $t$ in a  \ServerList of \system, there is no non-full server between $\head{v}$ and $\hostt{v}$.
\end{myInvariant}

\subsection{Access Operations}
\label{sec: list}

\begin{figure*}[t]
    \captionsetup[subfigure]{justification=centering}
    \begin{subfigure}[b]{0.24\textwidth}
    \centering
    \includegraphics[width=\textwidth, clip]{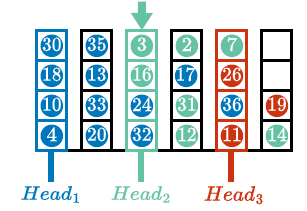}
    \caption{Access step.}
    \label{subfig: access}  
    \end{subfigure}
    \begin{subfigure}[b]{0.24\textwidth}
    \centering
    \includegraphics[width=\textwidth, clip]{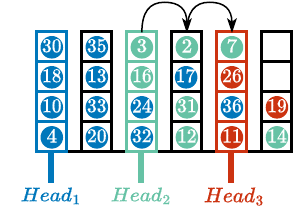}
    \caption{Search step.}
    \label{subfig: search}  
    \end{subfigure}
    \begin{subfigure}[b]{0.24\textwidth}
    \centering
    \includegraphics[width=\textwidth, clip]{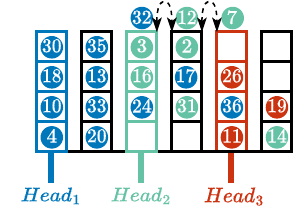}
    \caption{Reconfiguration step.}
    \label{subfig: reconfigure} 
    \end{subfigure}
    \begin{subfigure}[b]{0.24\textwidth}
    \centering
    \includegraphics[width=\textwidth, clip]{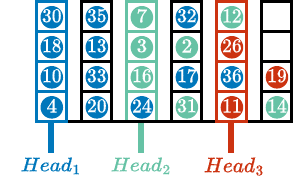}
    \caption{Final configuration.}
    \label{subfig: final} 
    \end{subfigure}
    \caption{An example of the \ServerList access problem with three heads and a capacity of four. In this figure, the relation between items and their original host is shown by using the same color.
    Assume access to the item $7$ with $Head_2$. 
    As it is a green item, that access starts from the $Head_2$, the head for green items (Figure~\ref{subfig: access}).
    Then we search server $Head_2$ and the servers that come after it for item $7$ (Figure~\ref{subfig: search}).
    After accessing item $7$ we swap it with the oldest items in servers between its current server and its host server (Figure~\ref{subfig: reconfigure}), to reach the final configuration (Figure~\ref{subfig: final}).
    }
    \label{fig: MTHC}
\end{figure*}

The traditional list access problem~\cite{SleatorT85} considers a list of servers (with capacity one) on a \emph{line}, and all items share the same head.
A \ServerList is an extension of a list in two directions:
\begin{itemize}
    \item We allow the list to have multiple heads, i.e., items can be assigned to one of the servers from a subset of $0 < b \le n$ servers, which we call heads of the \ServerList. We denote these heads by $\{head_1,\dots,head_b\}$\footnote{This definition generalizes definition needed to prove the competitiveness of \system, in which every server would be a head.}.

    \item We allow servers to have capacities larger than one and equal to $c$, and we consider all of the servers to be full except the last one. 
\end{itemize} 
In the following, we will show that a competitive result can be achieved for \ServerLists. 
The algorithm that we use for access in a \ServerList is the same as Algorithm~\ref{Alg: MTHC}. We can see the steps of the access algorithm on a \ServerList in Figure~\ref{fig: MTHC}.

In the \ServerList maintained by Algorithm~\ref{Alg: MTHC}, between two items with the same head, the item that has been accessed more recently is closer to the head. It is because after an access request, the least recently accessed item is moved to its head, and we have chosen the oldest items from servers in between to move back.
\begin{myInvariant}
\label{inv: recent}
For any given pair of items $u$ and $v$ with the same head $H$, if the last access to $u$ was sooner than $v$, item $u$ is closer to head $H$.
\end{myInvariant}

Our algorithm achieves a low access cost, and as we show in the following, a low competitive ratio. To show this, we compare the amortized total cost of our algorithm with the optimal offline algorithm \OPT over the whole request sequence. 
We consider the optimal algorithm that is offline, i.e., it is aware of the input sequence in advance, hence this proof shows that Algorithm~\ref{Alg: MTHC} performs almost optimal, even without any additional information about possible future requests or patterns in the input.
Furthermore, among such optimal offline algorithms, we select an algorithm that only reconfigures items between servers, i.e., keeps items in each server in the order that they have been added.
Such an optimal offline algorithm exists, as the order of items inside a server does not affect the cost of an algorithm, since only movements between servers are costly. Proof of the following theorem uses an argument based on counting the number of inversions between the optimal offline algorithm and our algorithm, to show competitiveness of our algorithm.\footnote{The proofs of statements marked by \appsymb{} are deferred to \Cref{app: omitted proofs}.}

\begin{mytheorem}{\appref{thm: MTHC}}
\label{thm: MTHC}
Algorithm~\ref{Alg: MTHC} is $2 \cdot (1+\omega)$ competitive, where $\omega$ is the cost of moving items between two adjacent servers. 
\end{mytheorem}
\appendixproof{thm: MTHC}
{
Before going into details of the proof, we define two concepts: \emph{position}, and \emph{inversion}.  
We first look at the position of an item inside the server that currently hosts it at time $t$. To define the position of an item, we focus on algorithms that maintain a pre-determined ordering of items inside all of their servers. We developed \system such that it has this property, and as discussed before, we consider such an \OPT algorithm.

\begin{mydefinition}[Inside position]
The inside position of an item $v$ at time $t$ is the number of items that are in front of it in the ordered list of its host, $\hostt{v}$.
\end{mydefinition}

The server that contains an item $v$ might change over time. This might be because the $\head{v}$ was full during the insertion of $v$, or due to reconfigurations of other items. In the following, we define the relative position for all items at any given time compared to any given head in a \ServerList.

\begin{mydefinition}[Headed relative position]
The headed relative position of an item $v$ that is in front of the head $H$ in an algorithm $ALG$ at time $t$, $\dist{ALG}{v, H}$, is its inside position at time $t$, plus $c$ times the number of servers between $\hostt{v}$ and $\head{v}$. 
\end{mydefinition}
The position for an item $v$ at time $t$ is its relative position to its head, i.e., $\dist{ALG}{v} = \dist{ALG}{v, \head{v}}$. 

We now define an \emph{asymmetric} inversion. We say there is an inversion between a pair of items $u$ and $v$ if they are in different orders in the \system and \OPT \ServerLists. 
The definition of inversion helps us determine ``how far" \system's \ServerList is from \OPT's \ServerList, which is key in comparing the cost of \system and \OPT.

\begin{mydefinition}[Inversion]
Considering an algorithm \OPT, there exists an asymmetric inversion between items $u$ and $v$ at time $t$ if and only if $u$ and $v$ are in front of $\head{u}$ and 
$\dist{\OPT}{v,\head{u}} < \dist{\OPT}{u}$ but $\dist{\system}{v,\head{u}}>\dist{\system}{u}$. 
\end{mydefinition}

As an example, let us consider \ServerList in Figure~\ref{subfig: access} as the OPT's \ServerList and the one in Figure~\ref{subfig: final} as \system's \ServerList. Then there exists an inversion between items $32$ and $7$, as item $32$ is closer to its head ($Head_1$) than $7$ in OPT's \ServerList, but vice versa in \system's \ServerList.

We emphasize that inversions are asymmetric, i.e., in one direction: $u$ might have an inversion with $v$ but not the other way around.
Furthermore, as we considered an \OPT algorithm that orders items inside a server as they have been moved to that server, similar to \system, ordering of items does inside a server not create an inversion.

We consider an indicator variable $\invt{u,v}$, which equals $1$ if there is an inversion between $u$ and $v$, $0$ otherwise.
Then we define the number of inversions in the \ServerList of \OPT at time $t$, as the sum of inversions for all items, and call it $\invtotalt{t}$. 
In other words, we have $\invtotalt{t} = \sum_{u,v \in V} \invt{u,v}$.

Using the above-mentioned definitions, next we show that our algorithm is constant competitive, using a potential function analysis for our model that considers a large enough request sequence.
\begin{proof}[Proof of ~\Cref{thm: MTHC}]
In the following proof, we consider the \ServerList of $OPT$ at time $t$, and see the effects of an access request at time $t+1$. 
To compute the amortized cost of \system, we consider the following potential function:
$\Phi^t = \frac{(1+\omega)}{c} \cdot \invtotalt{t}$. 
Consider $C^t_{\system}$ as the cost of system at time $t$, and $\Delta \Phi^{t \rightarrow t+1}$ as the change in potential from time $t$ to $t+1$.
We focus on the changes in the potential function, as \OPT and \system start from the same configuration, hence $\Phi^0 = 0$. Furthermore, given that the number of inversions is always non-negative ($\invtotalt{t} \ge 0, \forall t \ge 0$), the potential function is always larger than $0$ ($\Phi^t \ge 0, \forall t \ge 0$) by definition.

Then, the amortized cost of \system at time $t$ ( $A^t_{\system}$) is defined as the cost of \system at $t$ plus the change in the potential:
$
A^t_{\system} = C^t_{\system} + \Delta \Phi^{t \rightarrow t+1}
$.

Now, we discuss changes in the potential function, i.e., changes in the number of inversions. We divide these changes into two cases: right after $v$ was accessed in \system, and after possible changes by \OPT, and conclude by total changes in potential function.

\noindent \textbf{Right after $v$ was accessed.} 
We start by considering inversions that include $v$.
\begin{enumerate}
    \item \emph{Increase in the number of inversions.}
    Right after $v$ was accessed, its position in the \ServerList of \system is $0$, as we move $v$ to its head.
    At this point, some inversions might be created for items between $v$ and its head in \ServerList of \OPT. There are at most $\dist{OPT}{v}$ items between $v$ and its head in the \OPT \ServerList; hence the number of inversions created is at most $\dist{OPT}{v}$.
    \item \emph{Decrease in the number of inversions.}     On the other hand, all the previous inversions of the $v$ get destroyed. 
    Those items were before $v$ in the \system \ServerList(there are $\dist{\system}{v}$ items), and are positioned after $v$ in \OPT \ServerList(at most $\dist{OPT}{v}$ of those items can be positioned before $v$ in \OPT). Hence at least $\dist{ALG}{v}-\dist{OPT}{v}$  inversions are destroyed.
\end{enumerate}

As the ordering of other items does not change in \system's \ServerList, their relative position does not change (keep in mind that the position consists of \emph{inside position}). Therefore, inversions that do not include $v$ are not created nor removed.

\noindent \textbf{After possible changes by \OPT.}
Each reconfiguration of \OPT creates at most $c$ new inversions, besides some inversions that might be removed. 
This is because, at most, one item might move from one server to the other, and might cause inversion with all items that were in its previous server.
Assuming that \OPT has moved $r$ times, \OPT movements result in maximum $c \cdot r$ change in the potential function.

\noindent \textbf{Total changes in potential function.} Combining all cases, the total change in the potential function from time $t$ to $t+1$ is:
\[
\Delta \Phi^{t \rightarrow t+1} = \frac{(1+\omega)}{c} \cdot (\invtotalt{t+1}-\invtotalt{t}) \le 
\frac{(1+\omega)}{c} \cdot (\dist{OPT}{v}
 -  [\dist{\system}{v} - \dist{OPT}{v}] + c \cdot r)  = 
\]
\[
 \frac{(1+\omega)}{c} \cdot (  2 \cdot \dist{OPT}{v} -\dist{\system}{v} + c \cdot r )
\]

Considering Equation~\ref{eq: total cost} from Section~\ref{sec: model},
the total cost of access request to an item $v$ by \system ($C^t_{\system}$) is $\frac{(1+\omega)}{c} \cdot \dist{\system}{v}$. 
That is because:
\begin{itemize}
    \item \system immediately moves back the item $v$ from its host to its head (as described in Algorithm~\ref{Alg: MTHC}). Given Invariant~\ref{inv: recent}, the access cost of our algorithm equals its position at the \ServerList of \system $\frac{\dis{\system}{v}}{c}$,
    \item given that each reconfiguration cost $\omega$ times the access, the reconfiguration cost is $\omega \cdot \frac{\dis{\system}{v}}{c}$
\end{itemize}

Now we go back to amortized cost of \system at time $t$~($A^t_{\system}$):

\[
A^t_{\system} = C^t_{\system} + \Delta \Phi^{t \rightarrow t+1} \le
\frac{(1+\omega)}{c} \cdot \dist{\system}{v} + \frac{(1+\omega)}{c} \cdot (  2 \cdot \dist{OPT}{v} -\dist{\system}{v} + c \cdot r )=
\]
\[
\frac{(1+\omega)}{c} \cdot (2 \cdot \dist{OPT}{v} + c \cdot r)
\]

Then we compare the cost of \OPT with the amortized cost of the system.
We know the cost of \OPT is the access cost due to the position of $v$ in \OPT, i.e., $\frac{\dist{OPT}{v}}{c}$, plus the reconfiguration cost, $r \cdot \omega$, hence the total cost of \OPT at time $t$ ($C^t_{OPT}$) is $\frac{\dist{OPT}{v}}{c} + r \cdot \omega$.
Then we have:
\[
A^t_{\system} \le \frac{(1+\omega)}{c} \cdot (2 \cdot \dist{OPT}{v} + c \cdot r) < 2 \cdot \frac{(1+\omega)}{c} (\dist{OPT}{v} + c \cdot r) \le 
\]
\[
2 \cdot (1+\omega) (\frac{\dist{OPT}{v}}{c} + r \cdot \omega )
\le  2 \cdot (1+\omega) \cdot C^t_{OPT}
\]

The second to last step is possible as we assume $\omega \ge 1$. Given the last inequality, we can ensure that the Algorithm~\ref{Alg: MTHC} is $2 \cdot (1+\omega)$ competitive.

Given that the potential function is initially $0$ (both our algorithm and the optimal algorithm start from the same state) and is always positive, by summing up costs at each time step, we have:
\[ A_{\system} \le 2 \cdot (1+\omega) \cdot C_{OPT}\]
\end{proof}
}

\subsection{Insertion and Deletions}
Now we prove that \system is constant competitive even under insertions and deletions.

While our algorithm works in general, as usual in the literature~\cite{RevisitCH21,Albers03OnlineBook,borodin2005onlineBook}, for analytical purposes, we consider the following assumptions for any algorithm. This assumption is practically motivated.

\begin{myAssumption}
\label{assump: well-behaved}
Accesses are more frequent than insertions and deletions: in particular, we assume that each phase has at least length $\sum_{i=1}^{n-1} e^{-\frac{\alpha^2}{2m}(i+1)}$ (in which $\alpha$ is the additive capacity of each server).  We call such a request sequence \emph{well-behaved}.
\end{myAssumption}

In order to justify constant competitiveness of \system, we first analyze the maximum expected length among all \emph{maximal sequence of consecutive full servers} (by maximal we mean a sequence of servers that has non-full servers before and after), or a \MSCFS for short. In doing so, we first analyze the probability of having a \MSCFS of length at least $\ell$.

Let random variable $L$ be the length of a \MSCFS.
Now we compute the probability of $L \ge  \ell$, i.e., having a \MSCFS with length at least $\ell$.

\begin{myLemma}
\label{lemma: prob consecutive full}   
The probability of having a maximal sequence of consecutive full servers with length at least $\ell$ is less than or equal to $e^{-\frac{\alpha^2}{2 \cdot m} \cdot (\ell+1)^2}$.
\end{myLemma}
\begin{proof}
Consider $s$ to be the immediate full server after a non-full server $s^-$ (only such servers can be a start of a \MSCFS).
Now, let us consider $a_k$ as the number of items such that their head (the server with the closest hash in a clockwise manner) is the $k-1$ server after from $s$.

Having a \MSCFS with length $\ell$ from server $s$ requires all servers starting from $s$ should have been head of "enough" items. Formally, it is equivalent to having at least $k \cdot c$ items with their head in any of $1 \le k \le \ell$ consecutive servers starting from $s$ (remember that $c$ is the capacity of each server). 
In other words,
\begin{equation}
\label{eq: first}
P(L \ge \ell) = P(\forall 1 \le k \le \ell, \sum_{j=1}^{k} a_{j}\ge k\cdot c)\
\end{equation}

However, we only need a weaker assumption in this proof, which is $\sum_{k=1} ^{\ell} \sum_{j=1}^{k} a_j \ge  \sum_{k=1}^{\ell} k\cdot c$, therefore we can rewrite Inequality~\ref{eq: first} as:
\[P(L \ge \ell) \le P(\sum_{k=1}^{\ell} \sum_{j=1}^{k} a_{j} \ge \frac{\ell \cdot (\ell+1)}{2} \cdot c)\]

Now, to simplify the right-hand side, let us assign weight $\ell - k$ to all items that their head have distance $k$ from $s$.
To formalize this, let us consider $X_1,X_2,\dots X_m$ as random variables for all items. If the head of an item $v$ is not in the range of $\ell$ servers including and after $s$, we set $X_v$ equal to $0$. Otherwise, $X_v$ is $\ell$ minus distance between the head of $v$ and $s$. Hence, $0 \le X_v \le \ell$.

Let us then consider $X=\sum_{v \in V} X_v$.
By definition of our random variables, we have the following which essentially comes from a double counting method:
\[P(\sum_{k=1}^{\ell} \sum_{j=1}^{k} a_{j} \ge \frac{\ell \cdot (\ell+1) \cdot c}{2} ) = P(\sum_{v \in V} X_v \ge \frac{\ell \cdot (\ell+1) \cdot c}{2}  )=P(X\ge \frac{\ell \cdot (\ell+1) \cdot c}{2} )\]

To go one step further, we now use Hoeffding's inequality~\cite{hoeffding1963probability}. As the heads of items are selected uniformly at random, $X_v$ is independent from other items' random variables. To compute $\mu = E[X]$, using linearity of expectations, we have $E[X]=\sum_{v \in V} E[X_v]$. As \emph{in expectation} $\frac{1}{n}$ of items are assigned to a server (due to the uniformity of the hash function), we have: 
\[\sum_{v \in V} E[X_v] = \sum_{v \in V} \frac{\ell \cdot (\ell+1)}{2 \cdot n} = \frac{m \cdot \ell \cdot (\ell+1)}{2 \cdot n}\]

By setting $\tau=(c - \frac{m}{n}) \cdot \frac{\ell \cdot (\ell+1)}{2}$, and using Hoeffding's inequality, and as $\alpha \le (c - \frac{m}{n})$ , we have:
\[P(X\ge \frac{\ell \cdot (\ell+1) \cdot c}{2}  ) = P( X \ge \mu + \tau) \le e^{-\frac{2 \cdot ((c - \frac{m}{n}) \frac{ \ell \cdot (\ell+1)}{2})^2}{m\cdot \ell^2}} \le  e^{-\frac{\alpha^2}{2\cdot m}(\ell+1)^2}\]

Putting everything together, we have:
\[P(L \ge \ell) \le P(X\ge \frac{\ell \cdot (\ell+1) \cdot c}{2}) \le e^{-\frac{\alpha^2}{2\cdot m}(\ell+1)^2}\]
\end{proof}

Given Lemma~\ref{lemma: prob consecutive full}, we now compute the maximum length among all \MSCFSs.

\begin{myLemma}
\label{Lemma: expected consecutive full}
The expected maximum length among all of the maximal sequence of consecutive full servers is less than or equal to $\sum_{i=1}^{n-1} e^{-\frac{\alpha^2}{2m}(i+1)^2}$.
\end{myLemma}
\begin{proof}

Consider $L_{max}$ is a random variable indicating the length of the maximum \MSCFS, and $L$ is a random variable that determines the length of any arbitrary \MSCFS. Hence:
\[ E[L_{max}]=\sum_{i=1}^{n} i \cdot P(L_{max}=i) \le \sum_{i=1}^{n} i \cdot P(L=i) \] 

The above inequality is true because event $L_{max}=i$ means that we have an \MSCFS with length $i$ and the length of all other \MSCFS is less than equal to $i$.
We then replace $P(L=i)$ with $P(L\ge i)-P(L \ge i+1)$: 
\[E[L_{max}] \le \sum_{i=1} ^{n} i \cdot (P(L\ge i)-P(L\ge i+1))= \sum_{i=1} ^{n} i \cdot P(L\ge i) - \sum_{i=1} ^{n} (i-1) \cdot P(L\ge i) = \sum_{i=1}^{n-1} P(L \ge i)\]

The last step is true given the extra capacity per server, we can not have all servers full, hence $ P(L \ge n) $ and therefore $n \cdot P(L \ge n) $ is $0$. Now, based on Lemma~\ref{lemma: prob consecutive full} we get:
\[ E[L_{max}] \le \sum_{i=1} ^{n-1} P(L \ge i) \le  \sum_{i=1} ^{n-1}
e^{-\frac{\alpha^2}{2 \cdot m}(i+1)^2} 
\]

Hence, the expected maximum length of a \MSCFS is less than or equal to $\sum_{i=1}^{n-1} e^{-\frac{\alpha^2}{2m}(i+1)^2}$.
\end{proof}
% }

Recall that insertions and deletions require capacity changes at the end of the phase, and in the following, we first discuss the effects of such changes on the competitive ratio using the result of Lemma~\ref{Lemma: expected consecutive full}.

\begin{myLemma}{\appref{lem: cap}}
\label{lem: cap}
A capacity change operation at the end of each phase increases the competitive ratio by an additive constant, in expectation.
\end{myLemma}
\appendixproof{lem: cap}
{ 

\begin{proof}
    We start this proof by considering the case that the capacity of all servers only changes by one. In the end, we discuss the effect of changes by a higher amount (which might happen due to server insertion/deletion). 
    Recall that for a previously full $s$, we take the item to the next non-full server in case of increased capacity, or move the item to the next non-full server in case of decreased capacity. In the rest we focus on capacity increase, capacity decrease follows similarly (details at the end).
    
    Let us call the length if $j$-th \MSCFS $L_j$. What we want to compute is 
    $E[\sum \binom{L_j}{2}]$, since in a server list with length $L_j$, we need to move one item from the non-full server $L_j$ times to the head server, the item afterwards $L_j-1$ times to the server after head and so on.
    We know that 
    \[E[\sum \binom{L_j}{2}] = \frac{1}{2} \cdot  E[\sum L_j \cdot (L_j-1)] \le \frac{1}{2}  E[ \sum L_j^2]\]
    We know that $\sum L_j \le n$. Let us define $L_{max}$ as the length of \MSCFS with maximum length. We know $L_j \le L_{max}$ holds for all random variables $L_j $, hence we have:
    \[E[\sum L^2_j] \le E[L_{max} \cdot \sum L_j] \le E[L_{max} \cdot n] = n \cdot E[L_{max}]\]
    And we have the value $E(L_{max})=\sum_{i=1}^{n-1} e^{-\frac{\alpha^2}{2m}(i+1)^2}$ from ~\ref{Lemma: expected consecutive full}.
    On the other hand, we know that $OPT$ needs to maintain search property, so it needs to at least move $\frac{n}{2}$ items. Furthermore, \Cref{assump: well-behaved} about well-behaved request sequence states that insertion deletions happen only $\sum_{i=1}^{n-1} e^{-\frac{\alpha^2}{2m}(i+1)^2}=E(L_{max})$ times. Hence, the amortized cost of capacity increase is $\frac{ n \cdot E(L_{max})}{ \frac{n}{2} \cdot E(L_{max})}$, i.e. $2$ in expectation.
    
    In case of capacity decrease, we know there is always a non-full server in \system (due to the additive extra capacity, on top of the minimum capacity required). Hence, there is no wrap-around (i.e., we do not go over the server $s$ again).
    Furthermore, since $m>>n$, tbe capacity of the last server is increased enough to cover the incoming items.Therefore, the cost of the algorithm depends on the length of intervals of the full servers.
    In case of capacity change by more than one, the cost of $OPT$ also increases, hence the constant increase competitive ratio remains.
\end{proof}
}

Theorem~\ref{thm: main} builds upon Lemma~\ref{lem: cap} and Theorem~\ref{thm: MTHC}, given realistic assumptions on input, mentioned at the beginning of this section. This theorem shows, that in expectation, how we can conclude that the amortized cost of each operation is constant, given that access operations have a constant cost, and insertion and deletions are rare enough to match our assumption.

\begin{mytheorem}{\appref{thm: main}}
\label{thm: main}
Considering a well-behaved request sequence, \system is constant competitive, in expectation.
\end{mytheorem}
\appendixproof{thm: main}
{ 
\begin{proof}
We start this proof by looking back at the competitiveness of access operations \emph{given} item (or server) insertion (or deletion), and then focus on the competitiveness of those operations themselves.
We emphasize that we compare \system with an optimal offline algorithm that has the same capacity for all servers.
Furthermore, among many possible optimal offline algorithms, we consider that the optimal offline algorithm also respects Invariant~\ref{inv: between}, otherwise, its search mechanism would not be compatible with the standard search mechanism~\cite{MirrokniTZ18}.
%otherwise, it requires additional mechanisms to search for an item after observing a non-full server.

\noindent \textbf{Competitiveness of access operations.} 
We have concluded in Theorem~\ref{thm: MTHC} that an access operation inside a \ServerList is constant competitive. We now detail why that remains true considering items (or server) insertion (or deletion).

Given Invariant~\ref{inv: order} from the previous section, the relative order of items that were previously in the system remains the same after insertions or deletions; in other words, the least recently used items are closer to their own head in each \ServerList. Furthermore, the newly added item itself is always added at the end of a \ServerList; hence it does not affect the order of items in that or any given \ServerLists.

After an insertion or deletion, there is a possibility that a \ServerList splits (or a newly added server becomes its own server), or two \ServerLists might join together:
\begin{itemize}
    \item In case of a \ServerList splitting, there would be no item left of the first \ServerList in the other \ServerList. Hence, there would not be any shared inversions between the two split \ServerLists, i.e., there would not be an inversion that consists of an item in one \ServerList and another item in the other.
    Hence, our potential function which is based on inversions can be computed considering inversions of two \ServerLists separately.
    \item In case of two \ServerLists merging, as two \ServerLists lists did not have a shared item before, there were no shared inversions.
    Therefore, the potential function after the merge does not differ from when two \ServerLists were split, i.e., the sum of potential functions of previously split \ServerLists.
\end{itemize}
Therefore, in both cases, the arguments in Theorem~\ref{thm: MTHC} still hold and the constant competitiveness of \system holds.

\noindent \textbf{Competitiveness of other operations.} 
We discussed in Lemma~\ref{lem: cap} why capacity changes are competitive. 
As the further cost of item insertion (or deletion) is to move an item over a server list to the next non-full server (or bring the item from the non-full server in case of deletion), the extra cost depends on the length of consecutive full-server. Given the \Cref{assump: well-behaved}, we can offset this cost, since we assume insertion and deletion happens at a rate equal to the expected length of consecutive full servers.
%Considering the cost of item insertion or deletion, any optimal algorithm should search the whole \ServerList to check if the (new) item was in it before or not (our second assumption); hence the cost of our algorithm matches the cost of an optimal algorithm in those cases as well.

Server deletion and insertion do not introduce an extra cost themselves, but rather the movement of items is the costly part. In other words, we can consider a server insertion as a series of item deletions from other servers first and then item insertions to that server. Given that any $OPT$ algorithm needs to pay at least $1$ for each movement of items, we can simply apply what we discussed before for individual item insertion and deletions.
\end{proof}
}

\section{Experimental Evaluation}
\label{sec: implement}
In this section, we complement our theoretical analysis of \system (the main contribution of this paper) by providing first insights into the empirical performance of \system. 

We use our own benchmarking tool to compare consistent hashing algorithms in terms of cost and memory utilization and then show the effect of our parameters on them. 
Our tool can create a variety of input sequences, both based on real-world clicks datasets~\cite{dataset} and also synthetic inputs with various \emph{temporal localities}. Temporal locality measures the probability of an item being repeated consecutively. 

\begin{figure*}[t]
    \captionsetup[subfigure]{justification=centering}
    \centering
    \begin{subfigure}[b]{0.43\textwidth} \includegraphics[width=\linewidth, clip]{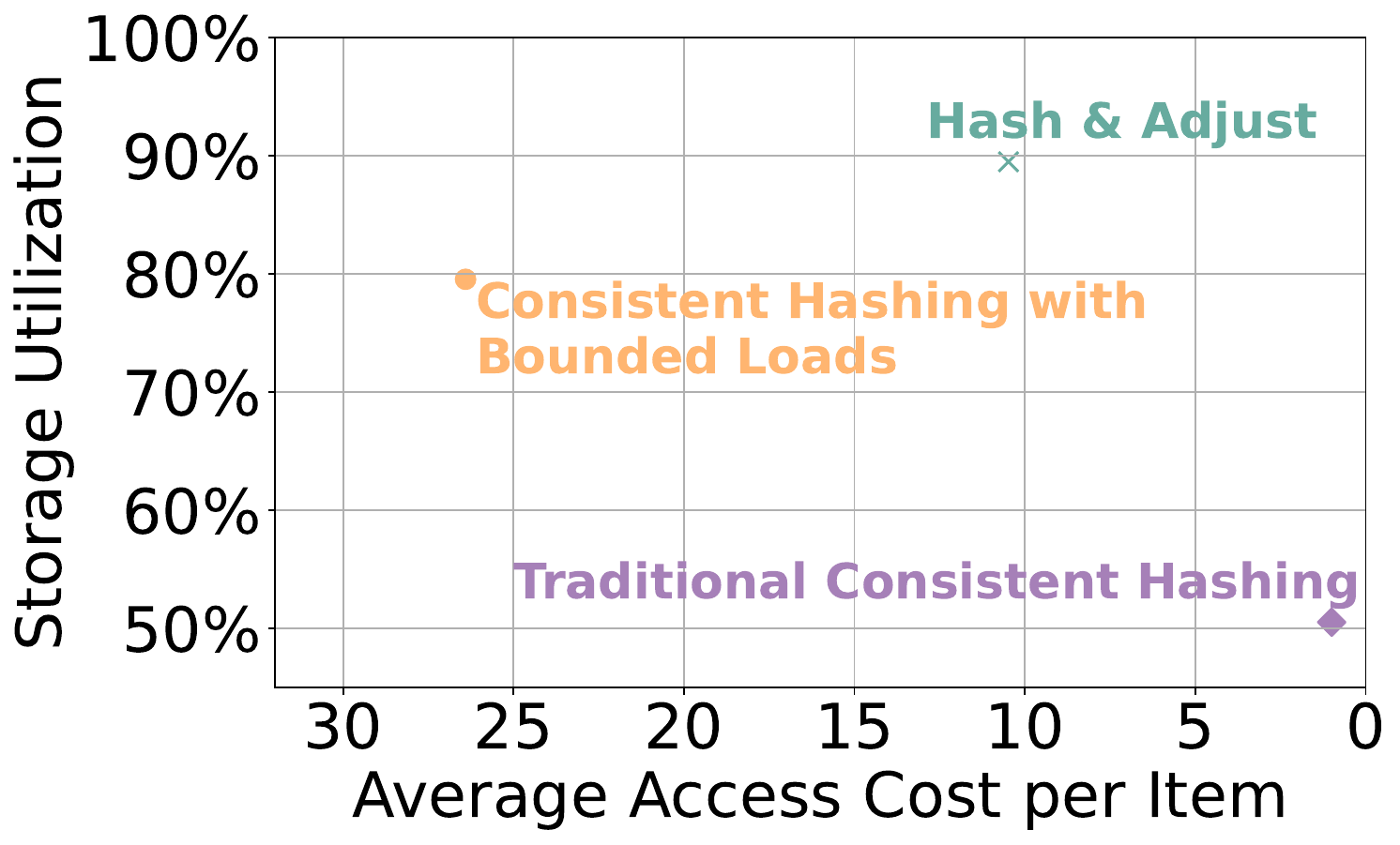}
    \caption{}
    \label{fig: intro-prac}  
    \end{subfigure}
    \begin{subfigure}[b]{0.39\textwidth}
    \centering
    \includegraphics[width=\textwidth, clip]{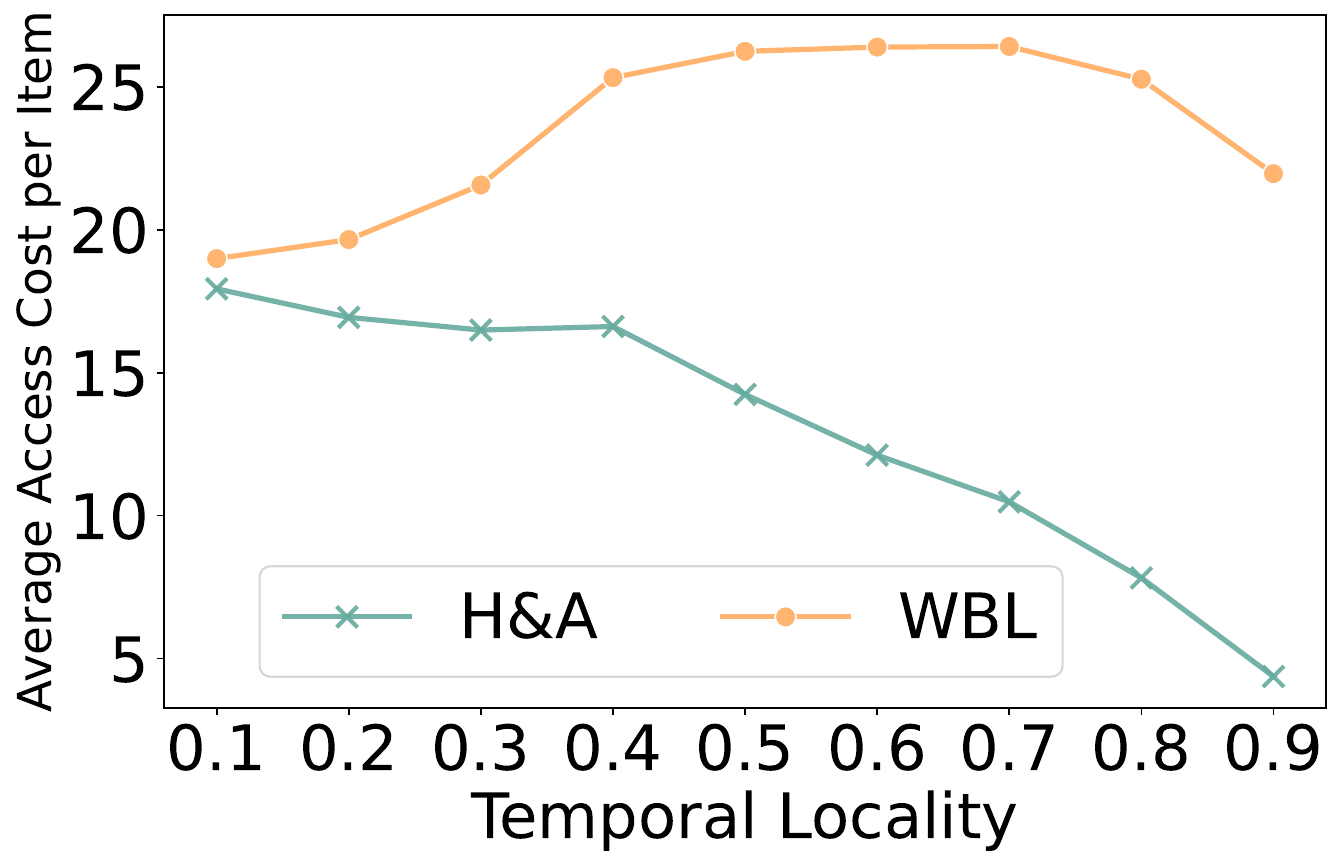}
    \caption{}
    \label{fig: res-loc-aug}  
    \end{subfigure}
    \caption{
    ~\Cref{fig: intro-prac} compares the average cost and memory utilization of the \system and the \WBL algorithm~\cite{MirrokniTZ18} and Traditional method~\cite{KargerClassic1}. We normalized the access cost by the number of items. This figure considers  $100,000$ requests, $10, 000$ items, and $20$ servers, and uses an instance generated by temporal locality $0.75$. Figure~\ref{fig: res-loc-aug} compares the access cost of \WBL with \system, by varying temporal locality of the input. For this figure, we consider same setup.%The traditional method has zero access cost, as it does not need to move items between servers because of its unlimited capacity. 
    }
    \label{fig: emp}
\end{figure*}

\subsection{Evaluation Setup}
Our experimental results are based Python~$3.6$ implementation.
For visualizations, we used seaborn~0.11~\cite{Waskom2021} and Matplotlib~3.5~\cite{Matplotlib} libraries.
The code was executed on a machine with 2x Intel Xeons E5-2697V3 SR1XF with 2.6 GHz, 14 cores each, and a total of 128 GB DDR4 RAM. A summary of parameters used in our benchmarking is in Table~\ref{table: in param} in the appendix.

\noindent \textbf{Input  generation.} 
We create a range of input sequences, both based on real-world  and also synthetic inputs. For real-world inputs, we considered a click dataset~\cite{dataset} and the CAIDA Anonymized Internet Traces Dataset~\cite{CAIDA}.

Our synthetic data set ranges over with various \emph{temporal localities}.
By varying temporal locality, we can adjust the number of consecutive repetitions of items that are assigned to a server. 
Such temporal bursts are a typical pattern in communication traffic
To formalize temporal locality, we consider the probability of an item being accessed consecutively. The temporal locality parameter has range of $[0.1,0.9]$, and has the value $0.7$ by default to simulate 
bursts in the input. We use $10,000$ and $1,000,000$ by default for the number of items and servers, respectively.

\noindent \textbf{Server insertion and deletion rate.} Given that we aim to utilize most of the servers' storage, we consider the hardware failure and replacement of servers to be independent of each other. As a result, in our evaluations, we used a Poisson distribution to simulate server (i.e., hard disk) insertion and deletion processes~\cite{YeXT13}.

\noindent  \textbf{Choice of hash function.}
The best choice of a hash function for the consistent hashing method has been a long long-held debate~\cite{AamandT19}. For our plots, we used more practically safe and secure SHA-512~\cite{DobraunigEM16Sha}. However, our implementation also supports 5-independent hash functions~\cite{K-independent1,5Independent1}.

\noindent  \textbf{Extra capacity.} 
Our algorithm, \system, and \WBL's algorithm rely on having extra capacity per each server. In our simulations, the default $\alpha$ value for our algorithm is $4$, i.e., four extra slots added to the minimum capacity of each server ($\lceil \frac{m}{n} \rceil$). In our evaluations, we considered the constant factor of $1.25$ for \WBL, based on their recommendation~\cite{MirrokniTZ18}.

\noindent  \textbf{Number of initial servers.}
To bootstrap an instance of consistent hashing, we need to set up a number of servers initially. We consider this value to be equal to $20$ by default.
    
\noindent \textbf{Stale time.} 
With \emph{stale time}, we aim to simulate the time-out option for each item stored in an instance of consistent hashing.
To do so, we delete an item after a fixed period of time, which has been set to $20$ minutes by default in our simulations.

\begin{figure}[t]
    \captionsetup[subfigure]{justification=centering}
    \centering
    \begin{subfigure}[b]{0.39\textwidth}
    \centering
    \includegraphics[width=\textwidth, clip]{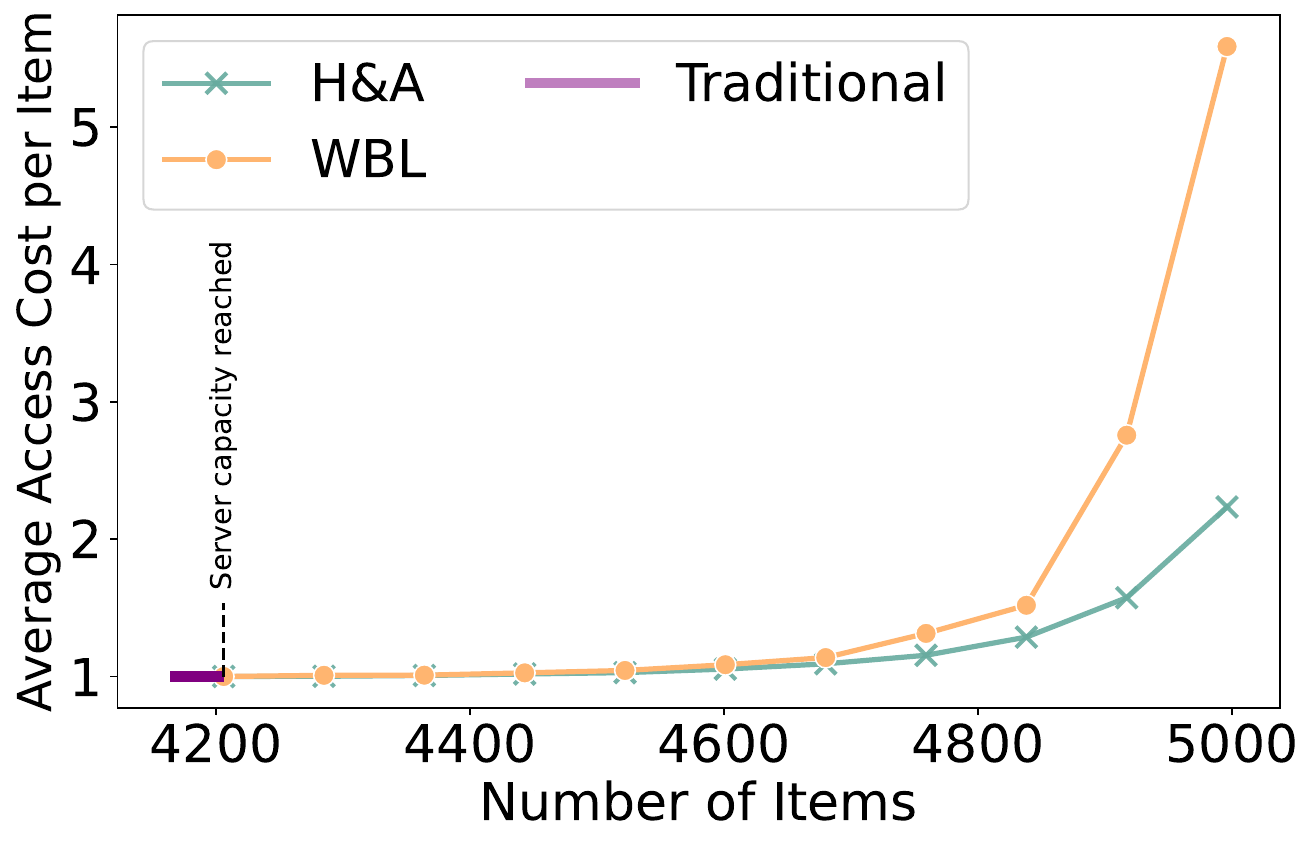}
    \caption{}
    \label{fig: sameCap1}  
    \end{subfigure}
    \begin{subfigure}[b]{0.41\textwidth}
    \centering
    \includegraphics[width=\textwidth, clip]{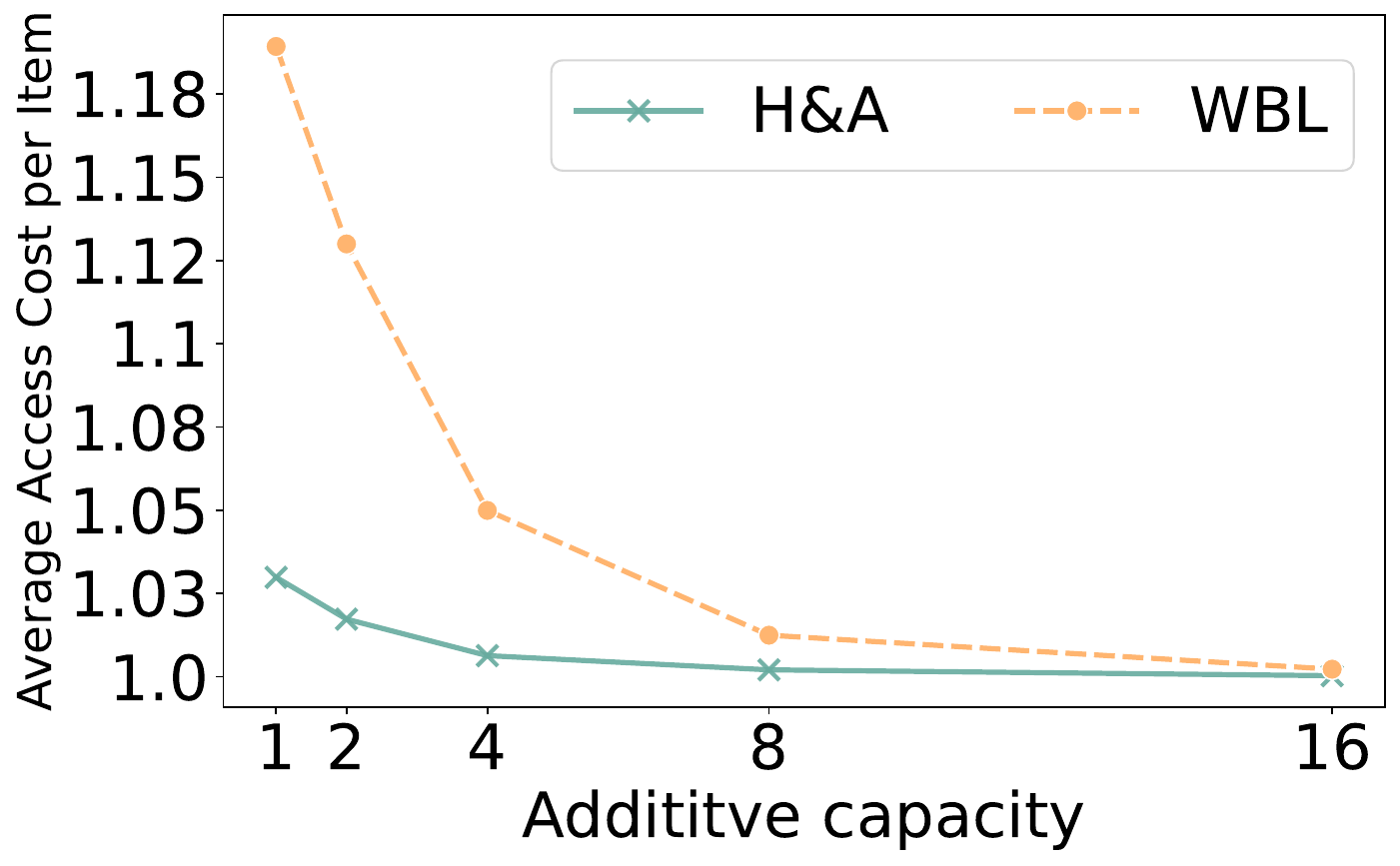}
    \caption{}
    \label{fig: sameCap2} 
    \end{subfigure}
    \caption{ Comparing average access cost of different algorithms considering the same capacity for all of the algorithms. Figure~\ref{fig: sameCap1} shows how the cost changes for an increasing number of items. This figure considers $50$ servers and $150,000$ requests. Figure~\ref{fig: sameCap1} shows how the cost changes for an increasing number of items. Here, the traditional algorithm stops when a server of it becomes full, and for other items, we run the experiment from scratch. The second figure considers $5,000$ items and the rest is similar to previous figure. Both figures consider the CAIDA~\cite{CAIDA} dataset.
    }
    \label{fig: emp2}
\end{figure}

\subsection{Experimental Results}

\noindent \textbf{Comparison of algorithms.} We compare our three main algorithms in terms of storage utilization and amortized access cost per request, given the real-world dataset.
Regarding storage utilization, using \WBL algorithm with a multiplication factor of $1.25$, the storage utilization only increases by $40$ percent from $50$ of the traditional algorithm (cf.~Figure~\ref{fig: intro-prac}). However, as also shown in the~figure, this better storage utilization comes with an access cost of $12$ per item on average.
Using \system, we can achieve closer to 100\% storage utilization, 90\% to be exact. Furthermore, we require more than 61\% less access cost compared to \WBL.

\noindent \textbf{Effect of temporal locality.} 
When we increase temporal locality (i.e., the probability of an item being requested consecutively), we can see a decrease in the cost of \system,  as shown in Figure~\ref{fig: res-loc-aug}.
That is because the item that has been moved to its head is going to be accessed again soon. The increase in the \WBL's access cost is because an item from further away might be accessed more often.

\noindent \textbf{Performance under same capacity.}
In this experiment, we consider that all algorithms have \emph{the same} capacity on the servers. As expected, \system performs much better when the additional capacity is small, as shown in Figure~\ref{fig: sameCap2}. Furthermore, for such a fixed capacity we can see our algorithm can tolerate more items with a lower cost, see Figure~\ref{fig: sameCap1}.

\begin{figure}[t]
    \captionsetup[subfigure]{justification=centering}
    \centering
    \begin{subfigure}[b]{0.39\textwidth}
    \centering
    \includegraphics[width=\textwidth, clip]{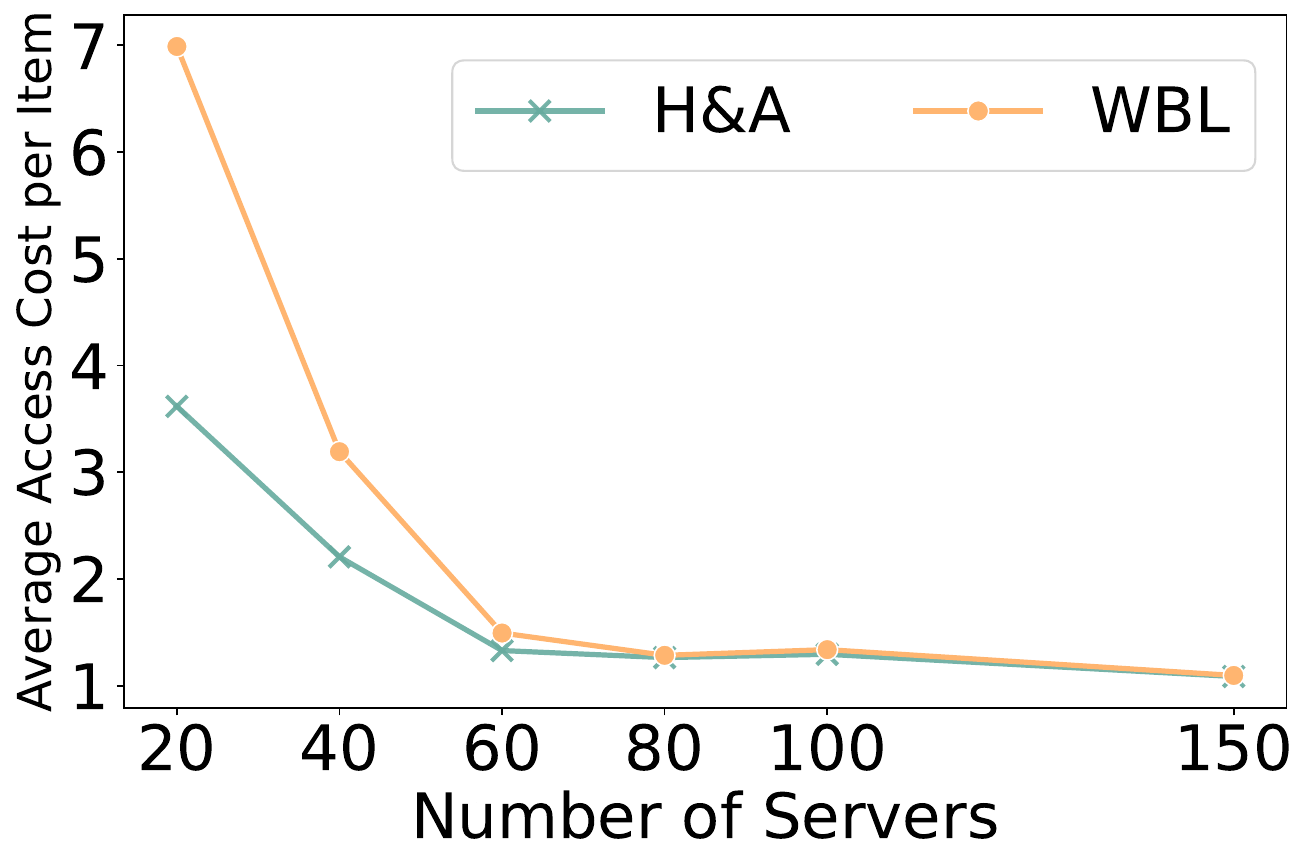}
    \caption{Changing number of servers.}
    \label{fig: res-intial}  
    \end{subfigure}
    \begin{subfigure}[b]{0.43\textwidth}
    \centering
    \includegraphics[width=\textwidth, clip]{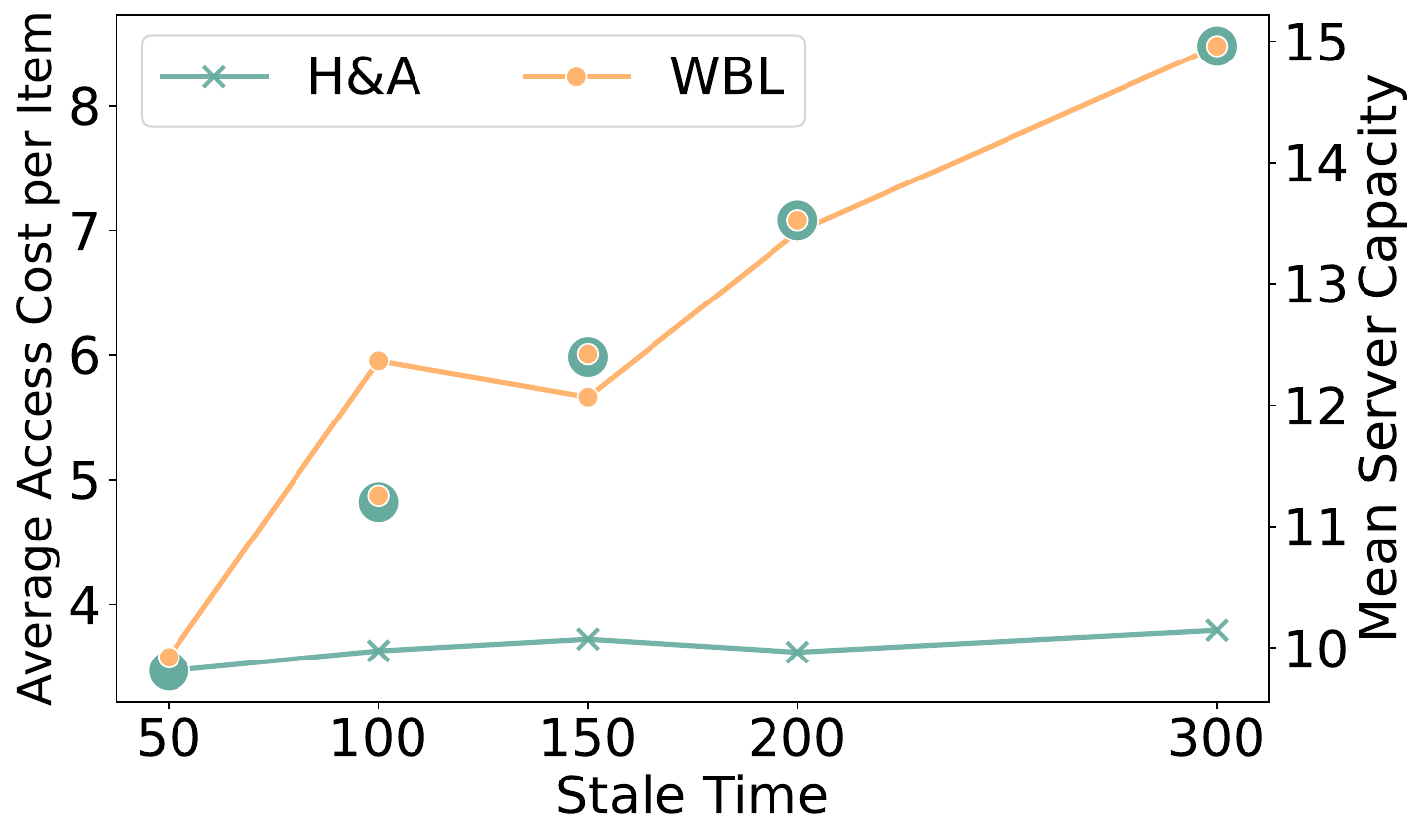}
    \caption{Changing stale time.}
    \label{fig: stale} 
    \end{subfigure}
    \caption{
    % Comparison of algorithms in terms of memory utilization and access cost, given the real-world dataset and $20$ servers. This figure is an empirical counterpart of Table~\ref{fig: intro}. The average access cost is adjusted by the number of items.
    Comparing the average cost of the \system and the "With Bounded Load" (\WBL) algorithm~\cite{MirrokniTZ18} based on two parameters of our benchmarking tool. We normalized the access cost by the number of items. The dots in Figure~\ref{fig: stale} shows the average server capacity by colored dots. First figure considers $1029$ items, and $100,000$ requests, and in second figure we consider $20$ servers. Both figures consider the click dataset~\cite{dataset}.} 
    \label{fig: emp3}
\end{figure}

\noindent \textbf{Effect of changing the number of servers.}
By increasing the number of servers (see Figure~\ref{fig: res-intial}), our algorithm and WBL both observe a reduced access cost. We can report from our observation that the sharp drop in access cost is because as we increase the number of servers, there is a higher chance that more non-full servers appear between a series of full servers, i.e., the maximum length of consecutive full servers decreases. This result is consistent with the outcome of Lemma~\ref{Lemma: expected consecutive full} that shows that the number of consecutive full servers decreases as the number of servers grows.

\noindent \textbf{Effect of other parameters.} We consider the effects of two further benchmarking parameters: the number of servers and stale time.
By increasing the number of servers (see Figure~\ref{fig: res-intial}), both algorithms observe a reduced access cost. We can report from our observation that the sharp drop in access cost is because as we increase the number of servers, there is a higher chance that more non-full servers appear between a series of full servers, i.e., the maximum length of consecutive full servers decreases.

Considering the stale time (Figure~\ref{fig: stale}), we saw an increase in the maximum capacity needed for servers, which was predictable, as more items will be in the system with a higher stale time.
However, the increase in the access cost of \WBL means that it performs worse in terms of both access cost and memory utilization for an increased stale time.

\section{Further Related Work}
\label{sec: related}
Our work builds upon literature in the context of consistent hashing and self-adjusting data structures. We review them in turn, showing how previous consistent hashing ideas result in increasing storage utilization, and how self-adjustments can lead to decreasing access time.

\noindent \textbf{Consistent hashing and storage utilization.} 
Consistent hashing was first introduced by Karger et al.~\cite{KargerClassic1}, building on top of the classical problem of ``balls into bins"~\cite{azar1994power}. The concept quickly became very important for distributed and peer-to-peer systems~\cite{DistrbutedBook20,LampingV14}.
The classic use case of consistent hashing is in the design of distributed hash tables in peer-to-peer systems like Chord~\cite{Chord03} or Kademlia~\cite{Kademlia02} and more~\cite{NaorW03,LuaCPSL05}.
The second use case of consistent hashing is in web caching~\cite{WebCaching99, RevisitCH21} in cloud systems (our experimental results explore this second scenario).

Recently, a team at Google implemented consistent hashing with bounded loads, showing  the effects of bounding load in their content hosting system~\cite{MirrokniTZ18}. Inspired by the work of Google, the team behind Viemo load balancer implemented a similar algorithm as part of HAProxy~\cite{tarreau2017haproxy} that resulted in 8-fold improvement in their cache bandwidth~\cite{rodland2016Vimeo}.

Traditionally, consistent hashing approaches abstract the routing protocol, i.e., allowing for flexible routing, as we considered in our paper.
Some of the related works suggest curtain approaches for routing. 
One of the proposals is using \emph{virtual servers}~\cite{Virtual01, Virtual04, SPAA04, AamandKT21, VirtualNodes2020}, in which a server is not responsible only for one hash value; but two or more.
On top of complicated routing, this method requires more storage to store the relation of each server with each hash function.
Another idea is to change the assignment of the next server for an item based on the number of full servers it observed so far~\cite{RevisitCH21}. This approach requires dedicated routing tables, and can only support insertion and deletion in specific scenarios. 

Furthermore, we point out two differences of our model with traditional caching models~\cite{CachingMarking,Roughgarden19}. 
Firstly, in such models cache hits come with zero cost, while in our model, accessing the first server has a cost of one.
Secondly, in our model, all servers provide the same speed in terms of accessing items, however, in the caching hierarchies, usually we see a slowdown when we get further away from processing units.

\noindent \textbf{Self-adjusting data structures and decreasing access time.}
Self-adjusting data structures have been explored for almost half a century now~\cite{CoffmanD73}.
The pioneering work of Sleator and Tarjan~\cite{SleatorT85} used the amortized analysis for the online algorithm of list update. 
To the best of our knowledge (check~\cite{INFOCOM23} for an updated list of papers), there are no results on the list update problem with capacity or multiple heads.

The notion of a self-adjusting hash table was first suggested by Pagli in 1985~\cite{Pagli85SelfAdjustingHash} 
and then by Wogulis~\cite{Wogulis89}, both as heuristics.
Self-adjusting hashing gained attention recently~\cite{LRUHashing18}, especially when Microsoft's team proposed VIP Hashing~\cite{VIPHashing}.

However, there are significant differences between our work, and the other suggested self-adjusting hash tables.
First of all, we emphasize that comparing the performance of our algorithm with methods of~\cite{Pagli85SelfAdjustingHash,Wogulis89, LRUHashing18, VIPHashing} is not possible: they aim to optimize \emph{intra}-server costs, but we aim to minimize the \emph{inter}-server costs.
On top of that, our work provides additional benefits such as:
(1) proving competitive guarantees formally, (2) balancing between storage utilization and access cost, and (3) analyzing the effects of temporal locality.
Furthermore, compared to VIP Hashing, our algorithm does not require a complicated learning phase and can be implemented with the change of a few lines in existing systems.

\section{Conclusion \& Future work}
\label{sec: conclusion}
Motivated by the dynamic temporal structure of demands in networked and distributed systems, we have introduced an adaptive approach to improve the performance of distributed hash tables.
Our main contribution is a constant-competitive algorithm for self-adjustments guaranteeing bounded loads.

In our future research, we aim to improve our approximation bounds and explore opportunities to render other distributed datastructures self-adjusting. 
From a practical point of view, we are working on incorporating our algorithms into real-world applications. In particular, we are working on an updated code for HAProxy, and also to improve load balancing for virtual IP address assignment (e.g., as proposed by Google~\cite{beyer2016site}).

\balance
\bibliographystyle{plainurl}
\bibliography{main}

\appendix

\newpage

\section{Omitted Proofs}
\label{app: omitted proofs}
In this section, we present the proofs omitted in the main body of the paper.
\appendixProofs

\section{Parameters of the benchmarking.}
In the Table~\ref{table: in param}, you can see a summary of the parameters that we used in our benchmarking.

\begin{table}[h]
\begin{center}
\begin{threeparttable}
\begin{tabular}{| c | c | c|}
   \hline
   Parameter  & 
   \cellcolor{gray!20} Range of values
    & Default
   \\ \hline
   \cellcolor{gray!20} Input generation &
   Click\tnote{1} or Temporal\tnote{2}  &
   Click
   \\ \hline
   Locality parameter &
   $[0.1,0.9]$ &
   $0.70$
   \\ \hline
   \cellcolor{gray!20} Number of Items (m) &
   $[10^3,10^9]$ items &
   $10,000$
   \\ \hline
   Number of Requests &
   $[10^4,10^{13}]$ requests &
   $1,000,000$
   \\ \hline
   \cellcolor{gray!20} Server insertion frequency&
   $[100,1000]$ minutes &
   $200$
   \\ \hline
    Server deletion frequency &
   $[100,1000]$ minutes &
   $200$
   \\ \hline
    \cellcolor{gray!20} Placement Algorithm &
   \system, WBL and Trad.\tnote{3} & \system
   \\ \hline
   Hash Function &
   $5$-Ind. or SHA-$512$ & SHA-$512$
   \\ \hline
    \cellcolor{gray!20}  Additive extra capacity &
   $[+1,+\lceil \frac{m}{n} \rceil]$ &
   $+4$
   \\ \hline
   Multiplicative extra capacity &
   $[1,2]$ &
   $+1.25$
   \\ \hline
    \cellcolor{gray!20} Number of Initial Servers (n)&
   $[20,10000]$ servers & $20$
   \\ \hline
   Stale time &
   $[20,200]$ minutes & $200$
   \\ \hline
\end{tabular}
\begin{tablenotes}
\item[1] Refers to the real-world click dataset~\cite{dataset}, 
\item[2] is the temporal locality generation method.
\item[3] "Traditional" method from~\cite{KargerClassic1}. 
\end{tablenotes}
\end{threeparttable}
\end{center}
\caption{Parameters and system that can be used in our benchmarking tool.}
\label{table: in param}
\end{table}

\end{document}